\documentclass[]{aa}

\usepackage[varg]{txfonts}
\usepackage{amsmath}
\usepackage{amsfonts}
\usepackage{amssymb}
\usepackage{graphicx}
\usepackage{natbib}
\bibpunct{(}{)}{;}{a}{}{,}

\begin{document}

\title{180\degr~Rotations in the Polarization Angle for Blazars}

\author{M.~H. Cohen\inst{1} \and T. Savolainen\inst{2,3,4}}

\institute{Department of Astronomy, California Institute of Technology, Pasadena, CA 91125, USA\\ \email{mhc@astro.caltech.edu}
   \and Aalto University Department of Electronics and Nanoengineering, PL 15500, FI-00076 Aalto, Finland\\ \email{tuomas.k.savolainen@aalto.fi}
   \and Aalto University Mets\"ahovi Radio Observatory, Mets\"ahovintie 114, FI-02540 Kylm\"al\"a, Finland
   \and Max-Planck-Institut f\"ur Radioastronomie, Auf dem H\"ugel 69, DE-53121 Bonn, Germany
}

\date{Received 22 October 2019; Accepted 13 March 2020}

                 %%%%   ABSTRACT   %%%%

%% A&A unstructured abstract. no refs, one paragraph
%\abstract{..}

%\begin{abstract}{..}
\abstract{
Rotations of the electric vector position angle (EVPA) in blazars are
often close to an integral multiple of 180\degr.  There are multiple
examples of this in the literature, and our analysis here, of the optical
polarization data from the RoboPol monitoring program, strengthens
the evidence by showing that $n\pi$ rotations occur more frequently
than expected by chance.  We explain this with a model consisting of
two polarized emission components: a ``jet'' that is constant in time,
and a ``burst'' that is variable. The EVPA of the combination is $\rm
EVPA_{jet}$ at both the beginning and the end of the burst, so the
net rotation across the burst must be $n\pi.$ Examples are analyzed on
the Stokes plane, where the winding number for the Stokes vector of the
combination gives the value of $n$.  The main conclusion is that the EVPA
rotation can be much larger than the physical rotation of the emission
region around the axis of the jet, but this requires the EVPAs of
the jet and the burst to be nearly orthogonal.  A shock-in-jet
calculation by Zhang et al. can provide a physical model for our
toy model, and in addition automatically gives the needed orthogonality.
The model is illustrated with data on OJ\,287 published by
Myserlis et al., and we suggest that the large rapid EVPA rotation
seen there might be a phase effect and not representative of a physical
rotation.} 

%\end{abstract}{..}


\maketitle

                 %%%  INTRODUCTION  %%%

%Section 1: Intro
\section{Introduction}
\label{s:intro}

Blazars typically show strongly variable emission at radio through X-ray
wavelengths, and some emit $\gamma$-rays and high-energy particles.
In this paper our interest is on the polarization of this emission
at radio, infrared, and optical wavelengths. The polarization is of
particular interest because it contains information on the geometry,
magnetic field structure and physical mechanism of the source, and
because large rapid polarization changes sometimes appear to be associated
with high-energy radiation.

The electric vector position angle (EVPA) can rotate substantially on a
short time scale; i.e., hours to days at optical to radio wavelengths,
respectively, although typically the scale is much longer.  The rotation
can persist for a year or more, and can be 360\degr~and larger.  The large
rotations have been discussed as tracers of a moving emission region
in a bent jet, or on a helical path, or in the disk \citep[e.g.,][]{K88,
ST92, M08, V10, MKA18}.  Linear models involving relativistic time-delay
effects have also been discussed \citep{B82, ZCB14, ZCB15}. Successive
large rotations with a reversal in the sense of rotation have been seen
\citep{CAA18}, with the reversal ascribed to MHD effects.

\citet{DM09} use three emission components to explain the behavior of
OJ\,287 at radio and near-IR wavelengths. The components are a narrow
fast spine, a slow sheath, and the boundary region between the spine and
the sheath. The variations in flux and polarization are due to transverse
motions of the jet, which cause variations in the viewing angle.

Stochastic models have also been used to explain the large rotations
\citep{JR85, J88, M14}. In this class of models the jet
is turbulent and contains a large number of cells whose magnetic fields
are oriented at random. At each time step the magnetic field in one cell
is changed at random, and the resulting random walk in the net EVPA
can occasionally include large rotations.  While many of the individual
rotations can be explained with such a stochastic model, it has been
shown that it is unlikely that \textit{all} the observed rotations would
be produced by a random walk mechanism \citep{BP15, KS16, KB17}.

In this paper we develop a two-component model consisting of a steady
``jet'' and a time-dependent ``burst''. The sum of these two can have a
large variety of EVPA rotations, some of which are presented below.
The general scheme of representing a polarization event as a sum of
two components is an old idea, \citep[e.g.,][]{B82, ST92, V10} and
the model we use is essentially the same as the one used our earlier
paper on OJ\,287 \citep{CAA18}.  However, in the present paper we
emphasize that the EVPA rotation can be a phase effect, and does
not necessarily represent a physical rotation of an emission region.
If the strengths of the two components become similar while the EVPAs
are nearly perpendicular, then the linear polarization of the sum will
have a minimum, while the EVPA of the sum rotates rapidly, up to nearly
180\degr. There is observational evidence for this effect, in that it
has often been noted that there is a polarization minimum at the time
of the fastest EVPA rotation. \citep[e.g.,][]{ST93, BP16a}.

A two-component model is also useful if the variability is a function of
frequency, in addition to time.  \citet{HBI84} used two components with
fixed EVPA and fractional polarization, and with variable flux density,
to explain the spectral polarization behavior of OJ\,287, including a
temporary 80\degr~change in EVPA as a function of frequency.

Two-component models have typically taken one component as being steady
while the other one is variable, and the combination produces the observed
changes \citep[e.g.,][]{CAA18}. The existence of a quasi-steady component
can be justified with observations.  Examples include a study of OJ\,287
at R-band by \citet{V10} who noted that there was an ``... underlying
stable source of polarized emission'' which they called the optical
polarization core (OPC). \citet{MKA18} similarily found a stable
EVPA for OJ\,287, at radio wavelengths.  \citet{PK17} showed, at radio
wavelengths, that while AGN generally are variable, 40\% of their sample
showed a tendency for a preferred EVPA in the VLBI core over time, while
\citet{JSE94} and \citet{HLB16} showed that many BL Lacs had a preferred
polarization angle.  However, there exists also a number of blazars that
show almost continuous variability in their optical EVPA \citep{MJ17}.

In this paper we present results on the Stokes (Q,U) plane, as loops
generated by the rotating Stokes vectors. The topology of the loops,
especially whether they do or do not enclose the origin of the (Q,U)
plane, controls whether or not the EVPA will have a large rotation.
The number of times a loop encloses the origin is called the index
or winding number of the loop \citep{A79}, and is positive if the
rotation is counter-clockwise (CCW), negative otherwise.  \citet{V10}
have discussed the EVPA of OJ\,287 at R-band in terms of these loops;
see also \citet{SUF11} for 1510--089 and \citet{LVR16} for CTA\,102.

The plan for this paper is as follows.  In Section~\ref{s:180rotations}
we review the observational evidence for EVPA rotations being
preferentially on the order of $n\pi$, where $n$ is an integer, and
then in Section~\ref{s:robopol} make this more precise by analyzing
the Robopol data \citep{BP15, BP16a}.  The result is that in this
data set, $n\pi$ rotations occur more often than expected by chance.
In Section~\ref{s:simple} we give a simple argument as to why an $n\pi$
rotation might be expected, when a source produces a burst
that adds to the steady emission from the jet.

In Section~\ref{s:2-comp} we describe our two-component model with the
aid of rotating vectors on the Stokes plane. Section~\ref{s:limits}
gives limits to the relative strength of the two components, and their
angular difference, for a 180\degr~rotation. Section~\ref{s:special}
shows some special cases where the geometry is constrained, but
the results are unexpected. In Section~\ref{s:noise} we show with
a simple model that weak fluctuations can generate 180\degr~swings
of the EVPA, provided the jet and the burst are nearly orthogonally
polarized. Section~\ref{s:physicalmodels} contains brief comments on two
theoretical ideas that might provide some support for the two-component
model.

The blazar OJ\.287 is used in Section~\ref{s:oj287} to illustrate some
of these ideas. Section~\ref{s:discussion} contains a discussion
of the results, and a summary is in Section~\ref{s:summary}.
Appendix~\ref{s:loops} contains a derivation of the limiting conditions
for a 180\degr~rotation for the two-component model.

%% SECTION 2

\section{180\degr~Rotations}
\label{s:180rotations}

EVPA rotations of the order of 180\degr~or~360\degr~have often been 
reported in the literature; e.g., for 
0727--115 \citep{AHA81}; 
OJ\,287 \citep{K88, CAA18};  
0954+658 \citep{MLT14};
3C\,279 \citep{SUF11, KS16}; 
1510--089 \citep{SUF11, BDA17};
BL\,Lac \citep{AHA81, ST93, RV13}; and
3C\,454.3 \citep{G17}, and
there are further examples from the RoboPol collaboration \citep{BP15,
BP16a, BP16b}.  The RoboPol group has carried out a large program of
optical polarization monitoring of a statistically well-defined sample
of blazars, and one of their aims is to study the statistics of the EVPA
rotation events. They report that the distribution of EVPA rotations peaks
near 180$^\circ$, although the peak is quite broad.  We now consider
this further and show that 180\degr~and~360\degr~rotations occur more
frequently in the RoboPol data than what is expected, if they come from
a uniform distribution.

%%%%%%%%%%%%%%%%%%%%%%%%%%%%%%%%%%%%%%%%%%%%%%%%%%%%%%

%Section 2.1
\subsection{Statistics of Observed 180\degr~Rotations from RoboPol}
\label{s:robopol}

\begin{table*}
\centering
\caption{Fully sampled EVPA rotations from the RoboPol program}
\label{Tab1}
\begin{tabular}{lrrrcl}
\hline\hline
Source          & $\Delta \theta_\mathrm{net}$ & $\sigma ( \Delta \theta_\mathrm{net})$ & $\lambda$   & $\Delta \theta_\mathrm{max}$ & References   \\
                &  (deg)                    &    (deg)                             & ($\sigma$)   &  (deg)                      &               \\ \hline
RBPLJ0045+2127  & \textbf{+188.8}             &   \textbf{15.1}                      & \textbf{0.6} &  +200                       & \citet{BP16b} \\
RBPLJ0136+4751  & -216.7                      &   8.3                                & 4.4          &  -114/-109\tablefootmark{a} & \citet{BP16b} \\
RBPLJ0721+7120  & \textbf{-180.3}             &   \textbf{7.7}                       & \textbf{0.1} &  -208                       & \citet{BP15}  \\
RBPLJ1512-0905  & \textbf{+177.1}             &   \textbf{21.3}                      & \textbf{0.1} &  +242                       & \citet{BP16a} \\
RBPLJ1555+1111\tablefootmark{b}  & +138.6     &    7.1                               & 5.8          &  +145                       & \citet{BP16a} \\
RBPLJ1558+5625  & +139.9                      &   15.9                               & 2.5          &  +222                       & \citet{BP15}  \\
RBPLJ1635+3808  & \textbf{-4.7}               &   \textbf{15.8}                      & \textbf{0.3} &  -119                       & \citet{BP16b} \\
RBPLJ1748+7005  & -207.5                      &   12.4                               & 2.2          &  -127                       & \citet{BP16a} \\
RBPLJ1751+0939  & \textbf{-197.7}             &   \textbf{9.1}                       & \textbf{1.9} &  -225                       & \citet{BP16b} \\
RBPLJ1800+7828  & \textbf{-166.6}             &   \textbf{11.9}                      & \textbf{1.1} &  -192                       & \citet{BP16a} \\
RBPLJ1806+6949  & \textbf{-361.6}             &   \textbf{8.4}                       & \textbf{0.2} &  -347                       & \citet{BP15}  \\
RBPLJ1836+3136  & \textbf{+12.5}              &   \textbf{18.0}                      & \textbf{0.7} &  +182                       & \citet{BP16b} \\
RBPLJ1927+6117  & -29.1                       &   8.2                                & 3.5          &  -105                       & \citet{BP15}  \\
RBPLJ2202+4216  & -153.7                      &   9.9                                & 2.7          &  -253                       & \citet{BP15}  \\   
RBPLJ2232+1143  & \textbf{-364.6}             &   \textbf{11.3}                      & \textbf{0.4} &  -312                       & \citet{BP15}  \\
                & \textbf{-170.5}             &   \textbf{5.8}                       & \textbf{1.6} &  -140                       & \citet{BP15}  \\
RBPLJ2243+2021  & \textbf{-168.8}             &   \textbf{9.3}                       & \textbf{1.2} &  -183                       & \citet{BP15}  \\
RBPLJ2253+1608  & -108.2                      &   5.9                                & 12.2         &  -129                       & \citet{BP15}  \\
                & +121.3                      &   7.8                                & 7.5          &  +145                       & \citet{BP16a} \\
\hline
\end{tabular}
\tablefoot{
  \tablefoottext{a}{The rotation is in two parts.}
  \tablefoottext{b}{First six data points were excluded.}
}
\end{table*}

The RoboPol optical polarization monitoring program is designed for
efficiently detecting EVPA rotations in statistically well-defined
samples of blazars \citep{PA14}. Here we use the RoboPol data to
examine the fraction of detected rotations that are consistent with
a net $n\pi$ change in the EVPA.

During the three years of monitoring, RoboPol detected 40 EVPA
rotation events in 24 sources \citep{BP15, BP16a, BP16b}. Their
definition of a ``rotation'' requires $>90$\degr~change in EVPA
consisting of at least four measurements with significant EVPA swings
between them. The detailed definition is given in \citet{BP15}. From
these 40 rotations, we have selected those in which there are at least
two measurements before and after the rotation event that can be used
to estimate the steady EVPA. However, there is variation in how
  steady the EVPA is outside of the rotation events. In order to
  remove cases in which no characterisc EVPA can be defined outside of
  the rotations, we have further filtered the data based on the
  scatter of the measurements. If the EVPA before or after a rotation
  event has (circular) standard deviation that is larger than 0.4
  times the (circular) standard deviation of a uniform distribution,
  we exclude that rotation from further analysis. The chosen threshold
  is arbitrary, but we note that the exact value does not
  significantly affect the results obtained here\footnote{We
      explored a range of thresholds from 0.3 to 0.6 and the
      overrepresentation of $n\pi$ EVPA rotations found in this paper
      is statistically significant ($p<0.05$) in all cases.}. The
  above selection leaves altogether 19 rotations in 17 sources, which
  are listed in Table~\ref{Tab1}.

Using RoboPol data, we have calculated the difference between the
weighted mean EVPA before and after the rotation event, $\Delta
\theta_\mathrm{net}$. If there is only one rotation for the given
  source, we include in $\Delta \theta_\mathrm{net}$ all data points
  except for those that belong to the rotation according to the
  RoboPol criteria. If there is more than one rotation in the EVPA
  curve, we include in $\Delta\theta_\mathrm{net}$ only those
  ``non-rotating'' data points that bracket the given rotation. In the
  case of RBPLJ1555+1111, we have also excluded the first six data points
  from the beginning of the measurement series, since their average
  EVPA clearly deviates from the average EVPA of the measurements
  closer to the rotation event. We calculate the variance of the
weighted mean EVPA as
\begin{equation}
\sigma^2_\mathrm{mean} = \frac{1}{\Sigma_i 1/\sigma'^2_i} ,
\end{equation}
where $\sigma'^2_i = \chi^2_\nu \sigma^2_i$. Here $\sigma_i$ is the
measurement error of $i$th EVPA point and $\chi^2_\nu$ is the reduced
chi-square for the weighted mean with $\nu$ degrees of freedom,
i.e., we scale the EVPA errors so that $\chi^2_\nu = 1$ for
$\sigma'^2_i$ in order to correct for potential
overdispersion. Finally, the error on $\Delta \theta_\mathrm{net}$ is
$\sigma(\Delta \theta_\mathrm{net}) = \sqrt{\sigma^2_\mathrm{mean,1} +
  \sigma^2_\mathrm{mean,2}}$, where $\sigma^2_\mathrm{mean,1}$ and
$\sigma^2_\mathrm{mean,2}$ are the errors of the mean EVPA before and
after the swing, respectively.  $\Delta \theta_\mathrm{net}$ and
$\sigma (\Delta \theta_\mathrm{net})$ are given in the columns 2 and 3
of Table~\ref{Tab1}.

In the fourth column of Table~\ref{Tab1} we show $\lambda$, the
minimum difference between $\Delta \theta_\mathrm{net}$ and $n \cdot
180^\circ$ in units of $\sigma$. The fraction of rotations that have
$\Delta \theta_\mathrm{net} = n \cdot 180^\circ$ within 2$\sigma$ is
high -- 11 out of 19 or 58\% (these are marked with a bold
font in Table~\ref{Tab1}). There are seven such rotations
consistent with $n=1$, two rotations consistent with $n=2$ and two
rotations consistent with $n=0$. The average uncertainty of $\Delta
\theta_\mathrm{net}$ is $\pm$11.0\degr, which means that we would on
average expect $(2 \cdot 2 \cdot 11.0) / 180 = 24$\% of the cases to
be consistent with $\Delta \theta_\mathrm{net} = n \cdot 180^\circ$
within 2$\sigma$, if the $\Delta \theta_\mathrm{net}$ were random and
uniformly distributed. The observed fraction of $n\pi$ rotations is
much higher than this. A simple Monte Carlo calculation gives a
probability of $p = 0.002$ for having 11 or more
rotations within 2$\sigma$ of $n\pi$, out of 19 cases in
total, if $\Delta \theta_\mathrm{net}$ are randomly drawn from a
uniform distribution between 0 and $n \cdot 180$\degr ($n \ne
0$). The $\sigma$ used in the Monte Carlo rounds are
bootstrapped from the observed values.

The original definition for RoboPol rotation events requires an EVPA
change of $>90^\circ$ \citep{BP15} and this may slightly bias our
statistics. If one only considers events with $|\Delta
\theta_\mathrm{net}| > 90^\circ$, there are 9 out of 16
rotations that are within 2$\sigma$ of $n\pi$. In this case a Monte
Carlo simulation gives a probability of $p = 0.006$ for
having 9 or more $n\pi$ events, out of 16 in total,
if they were drawn from a uniform distribution. Hence, we conclude
that $\Delta \theta_\mathrm{net}$ in the RoboPol data are unlikely to
be uniformly distributed and that there seems to be overrepresentation
of $n\pi$ rotations, which is consistent with the model we propose in
Section~\ref{s:simple}.

Table~\ref{Tab1} also shows $\Delta \theta_\mathrm{max}$, the length
of the continuous rotation determined in \citet{BP15, BP16a, BP16b}.
Out of the 11 $n\pi$ rotations marked in boldface in
Table~\ref{Tab1}, there are 5 cases in which $\Delta \theta_\mathrm{max}$
significantly ``overshoots'' $\Delta \theta_\mathrm{net}$.  Such a
behaviour is seen in Figures~\ref{Fig6}c and \ref{Fig7}c and
can be explained by a multicomponent model or a model including
internal noise-like variability in EVPA. In 3 cases out
of 19, $|\Delta \theta_\mathrm{max}|$ falls short of the
total change in the average EVPA before and after the rotation
event. This can be explained if part of the continuous rotation is
missed due to a gap in the sampling. Another cause for too short
measured $|\Delta \theta_\mathrm{max}|$ values comes from the
definition of the rotation used in RoboPol papers: the rotation is
terminated when there is a change of sign in the EVPA swing. This
means that $\Delta \theta_\mathrm{max}$ will miss non-monotonic
180\degr~rotation events such as the one shown in Figure~\ref{Fig7}c.

%%%%%%%%%%%%%%%%%%%%%%%%%%%%%%%%%%%%%%%

% SECTION 3
\section{A Simple Argument for 180\degr}
\label{s:simple}

A simple argument shows that 180\degr~rotations are expected in some
circumstances. Consider the combination of a steady emission component
and a temporary one that is burst-like; i.e. it rises from zero, goes
through a maximum and then subsides back to zero. At the beginning and
end of this event the EVPA of the sum is $\rm EVPA_{jet}$ and so the
change in EVPA across the burst is 0\degr. This becomes non-trivial and
interesting when we note that, for EVPA, 0\degr~is the same as $n\pi$
where $n$ is an integer, and so the net EVPA swing must be $n\pi$. When
the conditions are right $n=1$ and the rotation will be 180\degr.  In the
next Sections we describe a simple 2-component model of this type,
and find the limiting circumstances for a 180\degr~rotation.

%%%%%%%%%%%%%%%%%%%%%%%%%%%%%%%%%%%%%%%%%%%%%%%%%%%%%%%%

%Section 4
\section{Two-Component Model}
\label{s:2-comp}

\begin{figure}[t]  
\centering
\includegraphics[width=1.0\columnwidth]{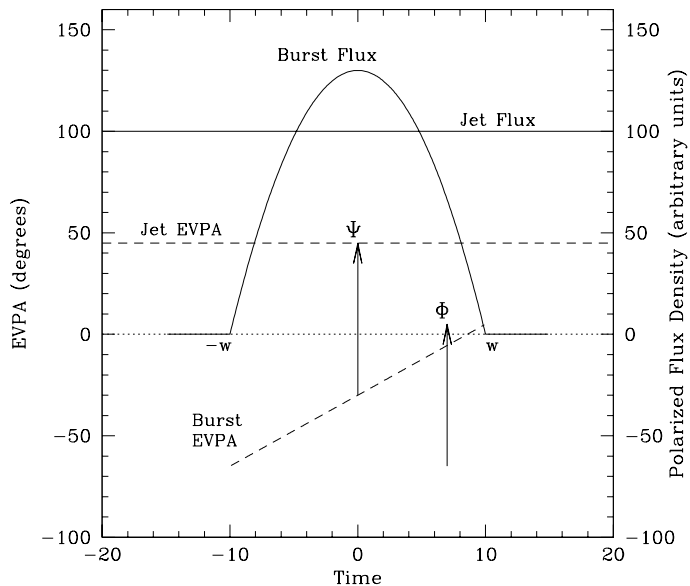}
\caption{
Two-component model. Linearly polarized flux density is shown
with solid lines, and EVPA is shown with dashed lines. Flux density
and time are in arbitrary units. $\Phi$ is the EVPA rotation across
the burst and $\Psi$ is the EVPA difference (jet -- burst) at time
$\rm t=0$. In this example $\Phi=70\degr$, $\Psi=75\degr$, and $\rm
R=ratio~of~peak~burst~flux~to~jet~flux =1.3$.
}
\label{Fig1}
\end{figure}

We use a simple model consisting of two components, a steady component
that we call the ``jet'' and a variable component called the ``burst''.
An example of the model is shown in Figure~\ref{Fig1}. This is the same
as the model used in \citet{CAA18}, see Figure~8, except that we now take
the burst to have a parabolic amplitude.  This burst is rather different
from the typical radio outburst, which often has a sharp rise followed by
a slower, approximately exponential, fall \citep[e.g.,][]{L84,VL99}. But
this shape applies to the total flux density, and we will be dealing
exclusively with the {\it polarized flux density}, which we denote
by PF, where P is the fractional linear polarization and F is the total
flux density.  Bursts in polarized flux can be highly irregular, because
subcomponents can partially cancel one another, even when the total flux
density is smooth. For simplicity, the burst has an EVPA that changes
linearly with time. This too is hardly realistic, and observed EVPAs
can be very irregular.

%%%%%%%%%%%%%%%%%%%%%%%%%%%%%%%%%%%%%%%%%%%  %%FIGURE 1
%%%%%%%%%%%%%%%%%%%%%%%%%%%%%%%%%%%%%%%%%%%

In Figure~\ref{Fig1} the jet parameters
are constant in time, with $\rm PF_{jet}=100$ and $\rm EVPA_{jet}=
45\degr$.  The burst has a parabolic shape with $\rm PF_{burst,max}=130$,
an EVPA that increases linearly in time, and PF and EVPA are zero
beyond $t=\pm \rm{w}  =\pm 10$. The important parameters are R, the flux
ratio, $\rm R=(PF_{burst,max}/PF_{jet})$; $\Phi$, the EVPA swing through
the burst, $\rm \Phi=(EVPA_{burst,w} - EVPA_{burst,-w})$; and $\Psi$,
the EVPA difference at t=0, $\rm \Psi = (EVPA_{jet} - EVPA_{burst,0})$.
In Figure~\ref{Fig1} $\rm R=1.3$, $\Phi=70\degr$, and $\Psi=75\degr$

%%%%%%%%%%%%%%%%%%%%%%%%%%%%%%%%%%%%%%%%%%%%%   FIGURE 2
%\begin{figure*}[pt]  
\begin{figure}[t]  
\centering
\includegraphics[width=1.0\columnwidth]{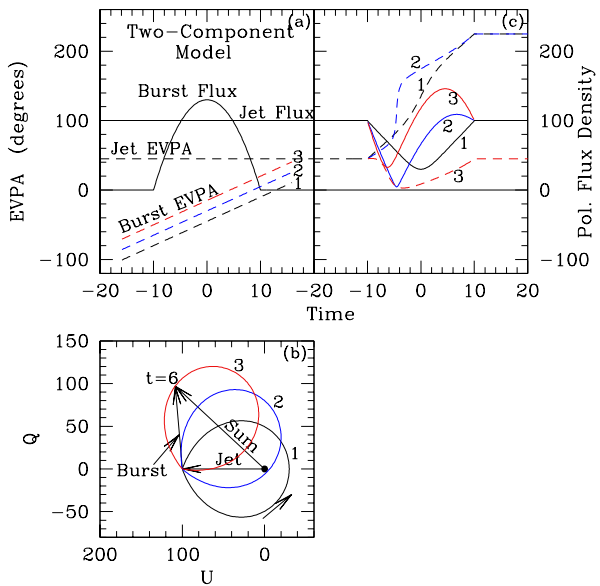} 
\caption{ 
(a) As in Figure~\ref{Fig1} but with three values of $\Psi$,
$\Psi_1=90\degr$, $\Psi_2=75\degr$ and $\Psi_3=60\degr$.  (b)
Vectors on the Stokes (Q,U) plane. The jet vector is stationary. The 3
burst vectors rotate CCW around the tip of the jet vector, starting at
$\rm t=-10$ and ending at $\rm t=+10$.  The sum vector rotates around
the origin.  The arrow at bottom right shows the direction of time.
The vectors for $\Psi_3$ are shown at $\rm t=+6$. Loops 1 and 2
enclose the origin but loop 3 does not.  (c) The polarized
flux density, PF, and the EVPA, for the 3 cases in (b). Note that cases
1 and 2, which enclose the origin in (b), have EVPA rotations of
180\degr, while case 3 has a mild swing of 42.0\degr~CW followed by
a return to the starting level. Note also in (c) that case 2 has a
deep PF minimum with a rapid EVPA swing, when the loop in (b)
gets close to the origin.  
} 
\label{Fig2} 
%\end{figure*} 
\end{figure}
%%%%%%%%%%%%%%%%%%%%%%%%%%%%%%%%%%%%%%%%%%%

The model in Figure~\ref{Fig2}a is the same as in Figure~\ref{Fig1}, 
but with three cases with different values of $\Psi$, $\Psi_1=90\degr$,
$\Psi_2=75\degr$ and $\Psi_3=60\degr$. The parameter $\Psi$ is important
for determining whether or not the EVPA can have a 180\degr~rotation.

The Stokes vectors for the models in Figure~\ref{Fig2}a are plotted on the
(Q,U) plane in Figure~\ref{Fig2}b.  The vector for the jet is
fixed with $\rm Q_{jet}=0$, $\rm U_{jet}=100$, and $\xi_{jet}=90\degr$,
where $\rm \xi = \arctan(U/Q) = 2 \cdot EVPA$.  The rotating vector
for the burst is added to the jet vector to form the sum vector, which
begins and ends its swing at the same place; namely, at the jet vector.
Thus the sum vector makes a closed loop. The three loops have the same
shape but their axes are rotated by $\Delta \xi = 2\Delta \Psi$.

Figure~\ref{Fig2}c shows the observable quantities, the time-dependent
linearly polarized flux density and the EVPA.  It is instructive to
examine the individual cases.  For cases 1 and 2 the loop encloses the
origin and so the total swing of $\rm \xi_{sum}$ is 360\degr, and the
total EVPA swing is $\xi_{sum}/2 = 180\degr$. Loop 2 comes close to the
origin of the Stokes plane, and at that time the sum vector becomes small
and its EVPA swings rapidly through about 90\degr, as can be seen in
Figure~\ref{Fig2}c. Loop 3 does not enclose the origin and from $t=-10$
to $t=-9.2$ it swings CCW, then it swings CW to $t=-3.3$, and then CCW to
$t=+10$; the total EVPA excursion is 42.5\degr~and the net EVPA rotation
is 0\degr.

The fluxes in Figure~\ref{Fig2}c can similarly be understood from the
loops in Figure~\ref{Fig2}b. Loop 1 is symmetric around the Q axis,
and $\rm PF_1$ is symmetric around $t=0$, where it has a minimum.
At that time the Stokes vectors for the jet and the burst are oppositely
directed; their EVPAs are perpendicular ($\Psi_1 = 90\degr$), and their
cancellation is maximum. Loop 2 comes close to the origin of the (Q,U)
plane, where the sum vector, i.e., the polarized flux density,
becomes very small. The sharp minimum in PF comes at the same time
as the peak rotation rate in EVPA.

That the peak rotation rate comes at the same time as the minimum in
PF (or P) is a general feature of two-component models, in which a
loop on the (Q,U) plane comes close to the origin. There is observational
evidence to support this feature.  For example, \citet{BP16a} state
that, in blazars, the polarization fraction P often has a minimum when
the EVPA is rotating most rapidly.  Further examples are given in e.g.,
\citet{ST93, M08, AAA10b, CAA18}.

\begin{figure}[ht]  %%FIGURE 3
\centering
%% left, bottom, right, top
\includegraphics[trim={0cm 8cm 0cm 0cm},clip,width=1.0\columnwidth]{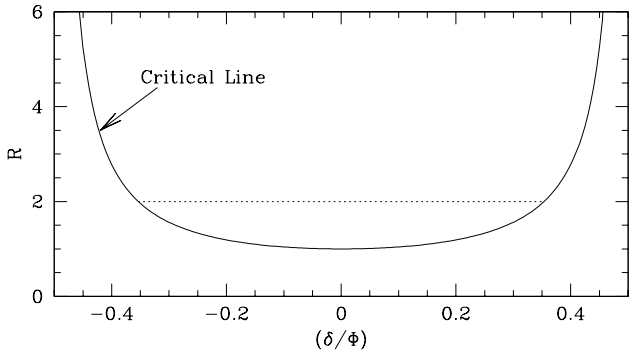}
\caption{
Critical line for a 180\degr~rotation, for the model in
Figure~\ref{Fig1}. The quantities are defined in Figure~\ref{Fig1},
and $\delta \equiv(\Psi-90\degr)$.  The critical line is the locus
of events ($\rm R, \Phi, \delta$) whose loops on the (Q,U) plane
touch the origin.  An event above the line has a loop that encloses
the origin and the EVPA rotation is 180\degr.  An event below the
line has a loop that does not enclose the origin, and its net EVPA
rotation is 0\degr.  The dotted line shows the range of ($\delta /
\Phi$) for a 180\degr~rotation, for $\rm R=2$. See text.
}
\label{Fig3}
\end{figure}

%Section 5. 
\section{Limits for the 180\degr~Rotation}
\label{s:limits}

In Figure~\ref{Fig2}b the nature of the EVPA curve (the winding
number) changes when the loop moves across the origin as $\Psi$
changes.  Similar effects are found as R and $\Phi$ vary. This is
summarized in Figure~\ref{Fig3}, which shows the ``critical line''
where the winding number changes from 0 below the line, to 1 above it.
The equation for this line is $\rm{R} = 1/(1-(2\delta/\Phi)^2)$ where
$\delta = (\Psi - 90\degr)$; i.e., $\delta$ is the departure from
orthogonality of the EVPAs of the jet and the burst. The critical line
is derived in the appendix.

Points on the critical line produce (Q,U) loops that touch the 
origin. Points above the line produce loops that enclose the origin
and have an EVPA rotation of 180\degr. Points below the line produce
loops that do not enclose the origin and have a net rotation of 0\degr.

In Figure~\ref{Fig3} the horizontal line at $\rm R=2$ shows
the allowed range of $\delta/\Phi$ for a 180\degr~ rotation. If
$\Phi$ is small, e.g., because the burst source is moving nearly
along the axis of the jet, then there still is the possibility of
a 180\degr~rotation if $\delta$ is small enough. For example, if
$\rm R=2$, there will be a 180\degr~rotation if
$|\delta|<\Phi/(2\sqrt{2})$.

If the jet and burst are independent, then $\delta$ can be regarded
as random, and for a fixed $\rm R~and~\Phi$ the probability of a
burst having a 180\degr~rotation is proportional to the length of its
horizontal line in Figure~\ref{Fig3}. For $\rm R=2~and~\Phi=10\degr$,
$|\delta| < 3.5\degr$ for a 180\degr~ rotation, and if $\delta$ is random
the probability for this is $7/180=0.04$.

Figure~\ref{Fig3} is valid only for our particular model, a parabolic
burst with a linearly-changing phase, superimposed on a steady jet.
Presumeably, a similar but more realistic model would have an analogous
line.

\begin{figure}[ht]  %%FIGURE 4
\centering
\includegraphics [width=1.0\columnwidth]{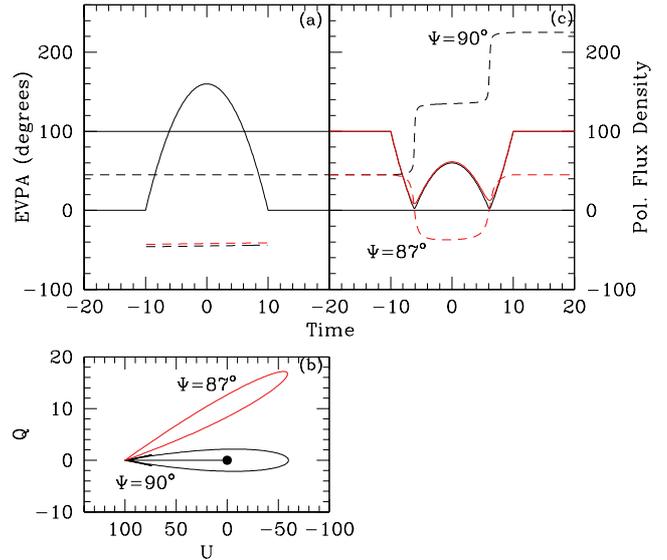}
\caption{
Rotations with $\Phi=2\degr$. Black: $\Psi=90\degr$
and the loop in (b) encloses the origin.  In (c) the two steps in
EVPA are in the same direction and the net rotation is 180\degr.  Red:
$\Psi=87\degr$ and the loop in (b) does not enclose the origin. In
(c) the two steps are in opposite directions and the net
rotation is 0\degr.  Note that the Q and U scales are different in
(b).  
}
\label{Fig4}
\end{figure}

\begin{figure}[ht]  %%FIGURE 5
\centering
\includegraphics[trim={0cm 1cm 1cm 0cm}, clip, width=1.0\columnwidth]{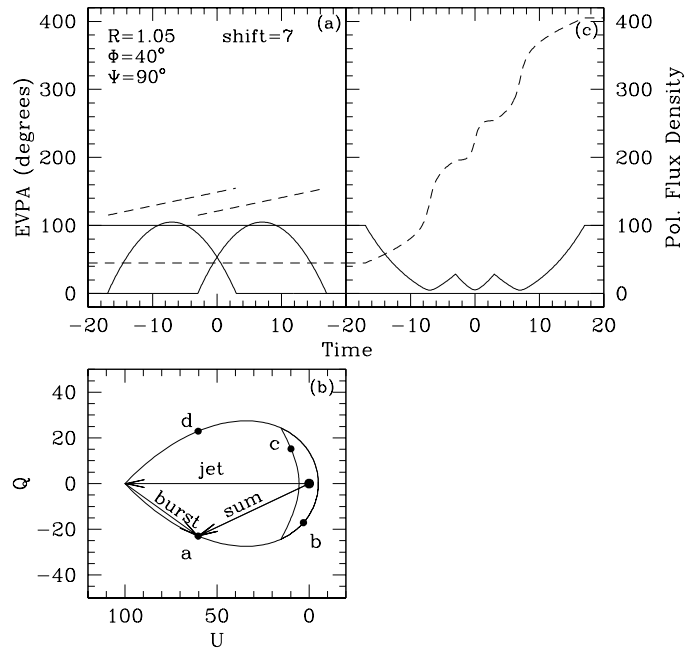}
\caption{
Model containing 2 bursts. (a) As in Figure 2a but with two bursts
shifted by $\rm \Delta t =\pm 7$. (b) as in Figure 2b. The time sequence
is a, b, c, b again, d. The loop encloses the origin twice, in the
CCW direction, so the EVPA rotation in (c) is 360\degr~CCW. See text.
}
\label{Fig5}
\end{figure}

%%%   SECTION 5.1
\subsection{Special Cases}
\label{s:special}

From the preceeding discussion, we see that even as $\Phi$
becomes small, there remains a finite range of $\Psi$, centered
on $\Psi=90\degr$, where the model can have an EVPA rotation of
180\degr. Figure~\ref{Fig4} shows the situation for $\Phi = 2\degr$
and two values for $\rm \Psi, 90\degr~and~87\degr$.  The loops in
Figure~\ref{Fig4}b are thin and the sum vectors sweep rapidly past
the origin, twice, putting $\rm PF_{sum}$ into the sharp minima seen
in Figure~\ref{Fig4}c. In Figure~\ref{Fig4}c the EVPA has two
separated steps of about 90\degr~each.  For $\Psi = 90\degr$ the steps
are both CCW and the total swing is 180\degr. But for $\Psi=87\degr$
the loop does not enclose the origin and the EVPA steps are in the
opposite sense, giving a net rotation of 0\degr. 

It may seem seem rather remarkable that such a small physical rotation,
2\degr~in this example, can lead to a 180\degr~swing in the observed
EVPA. We noted the possibility of such differences in our earlier paper
on OJ\,287 \citep{CAA18}, but without developing the general picture that
we have here. The effect is a consequence of the circumstances discussed
in Section~\ref{s:simple}. The EVPA across the burst changes by $n\pi$,
and in Figure~\ref{Fig4}, $n=0$ (red) and $n=1$ (black). This requires
a carefully chosen set of parameters, and in particular $\Psi$
must be close to $90\degr$. As seen in Figure~\ref{Fig3} the probability
for this is small, if $\Psi$ is random.  However, it seems unlikely that
$\Psi$ is random, and $\Psi \sim 90\degr$ might not be rare in blazars.

\citet{LPG05} have shown that, for an unresolved optically thin jet
with a helical magnetic field, the jet polarization should be either
along or across the jet axis. Observational examples of this include
\citet{HLA18}, who showed for BL Lacs that the core polarization is
preferentially along the jet. \citet{DM09} showed that in 2005 at 43 GHz
the polarization of an emerging new component in OJ\,287 was along the
jet while the polarization of the background jet was perpendicular to
the jet. See also \citet[][Figure 8]{AAA10a}, where the blazar 1502+106
is seen at 15 GHz to have core EVPA roughly perpendicular to the jet
in 2007, and parallel to the jet in 2008; and \citet[][Figure 6]{AAA16}
where IES 1011+496 similarly has core polarization roughly perpendicular
to the jet in 2010, and along the jet in 2012. In both these last two
cases there were gamma-ray flares during the polarization rotation.

The double 90\degr~step seen in Figure~\ref{Fig4}c is closely analogous
to the similar double 90\degr~step in the relativistic shock-in-jet
calculation by \citet[][see Figures 7 and 17]{ZCB14}. In both
cases the second step comes when the disturbance relaxes and the
system reverts to its original state. This is discussed further in
Section~\ref{s:discussion}.

The value for $\Phi$ used in Figure~\ref{Fig4}, $\Phi = 2\degr$, is well
below the fluctuation level seen in real situations, and so we expect
that the topology of the loop, and the EVPA curve, could be strongly
affected by fluctuations in the received signal.  This is considered
in Section~\ref{s:noise}.

Real bursts are generally not smooth like the parabola in Figure~\ref{Fig1}
and to simulate that we now allow the burst to have subcomponents.
Figure~\ref{Fig5} shows a case where the burst consists
of two parabolas like that in Figure~\ref{Fig1}, with peaks at $t=\pm
7$. The EVPAs of the two subcomponents are similarly shifted.  On the
Stokes plane, Figure~\ref{Fig5}b, the sum vector rotates CCW, successively
passing the points a, b, c, b again, and d. It makes two complete
loops around the origin, and the EVPA rotates by 360\degr, as shown in
Figure~\ref{Fig5}c.

\begin{figure}[ht]  %%FIGURE 6
\centering
\includegraphics[trim={0cm 1.5cm 1cm 1cm}, clip, width=1.0\columnwidth]{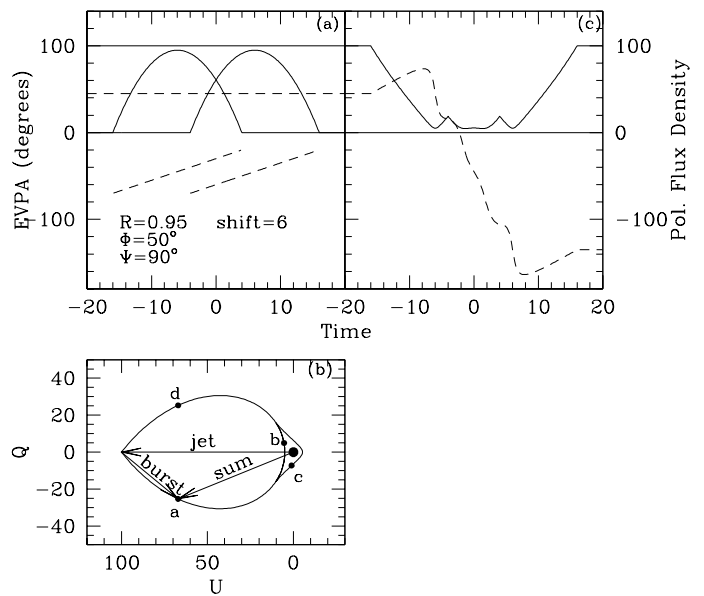}
\caption{
(a) As in Figure 5 but with $\rm R=0.95$. (b) The time sequence is
a, b, c, b again, d. The small loop encloses
the origin once in the CW direction, and in (c) the net EVPA rotation
is 180\degr, although the total swing is 237.4\degr. Note in (c) that the
rotation is CW although in (a) the rotation of the bursts is CCW.
Note also that the Q and U scales are different in (b).
}
\label{Fig6}
\end{figure}

The parameter R must be tightly tuned for the 2--loop situation seen
in Figure~\ref{Fig5}, where $\rm R=1.05$. If R is reduced below 1.0
the topology changes and the sum vector rotates around the small
loop in the opposite direction. Figure~\ref{Fig6} shows an example
of this behavior. The subcomponents are as in Figure~\ref{Fig5}, but
with $\rm R=0.95$. The sum vector successively passes the points a,
b, c, b again, and d, and the direction of rotation around the small
loop is CW. The total swing, shown in Figure~\ref{Fig6}c, is 237\degr,
i.e., larger than 180\degr, and the general rotation direction is CW,
even though the rotation direction for the subcomponents is CCW.

The three cases shown in this Section are special and require closely
adjusted parameters.  However, they do show that a simple interpretation
of an observed EVPA rotation, in terms of e.g., an emission region
moving on a helical trajectory, might be wrong.  The EVPA rotation can
be a phase effect; it can be much larger than the physical rotation, and
can even be in the opposite direction.  
Furthermore, a burst with
a single sense of rotation can produce EVPA rotations in both senses,
as in Figure~\ref{Fig4}c. This is particularly interesting because the
observation of both rotation senses has been taken as an indication
of a stochastic process \citep{KS16}, although here it results from a
deterministic event.  All these possibilities should be kept in mind
when EVPA rotations are interpreted in terms of both physical and
stochastic models.

\begin{figure}[t]  %% FIGURE 7 
\centering
\includegraphics[trim={0cm 1.5cm 1cm 1cm}, clip, width=1.0\columnwidth]{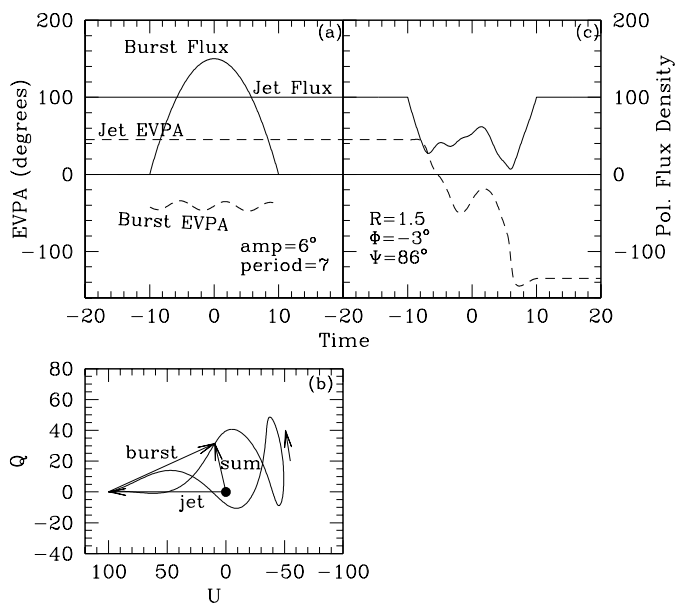}
\caption{
Two--component model that simulates a burst with a fluctuating EVPA. (a) The
burst EVPA has an overall slope of $-3\degr$ and is modulated by a sine
function with amplitude 6\degr. (b) The sum vector makes three loops on the
(Q,U) plane, with the middle loop enclosing the origin. The jet, burst
and sum vectors are shown at $\rm t=-6$. The arrow shows the direction
of time.  In (c) the EVPA shows a non-monotonic CW swing of 180\degr~plus
a small overshoot.  Note that the Q and U scales in (b) are different.
}
\label{Fig7}
\end{figure}

%Section 6 
\section{Source Fluctuations and Measurement Noise} 
\label{s:noise}

In our model the variable component changes smoothly, but in the real
world the emission has rapid changes in PF and EVPA. As we
now show, even small fluctuations can make large changes in the EVPA.

To simulate the effect of small fluctuations in the emission, we
keep the two-component model with a steady jet, but allow the EVPA
of the burst to have a sinusoidal ripple. This introduces 3 more
parameters, the amplitude, period, and phase of the sine function,
and the possibilities are thereby expanded. Figure~\ref{Fig7} shows an
example. In Figure~\ref{Fig7}, $\rm R=1.5, \Psi=86\degr$ and the EVPA
of the burst has a ripple of amplitude 6\degr, with an overall slope of
$\Phi=-3\degr$.  The motion of the sum vector defines 3 loops, with
the central one enclosing the origin and generating a CW EVPA rotation
of 180\degr, as seen in Figure~\ref{Fig7}c. In this case the topology in
Figure~\ref{Fig7}b is set by the sine wave and the overall slope $\Phi$
could be changed from --3\degr~to +3\degr~ with little effect. If R is
increased the pattern expands outward and when the origin is not in a loop
there is not a 180\degr~rotation across the burst. But when the expansion
is continued the innermost loop encloses the origin and again there is a
180\degr~rotation, but in the CCW sense. In this case the 180\degr~would
consist of two 90\degr~CCW swings, near the beginning and end of the
burst. If R is decreased below unity there are no 180\degr~rotations. 

In some circumstances, measurement noise can also be responsible for
large changes in EVPA. We consider an example of this with reference to
Figure~\ref{Fig2}. In Figure~\ref{Fig2}b the loop for case 2 comes
close to the origin of the (Q,U) plane, and at that time the amplitude
of the sum vector (i.e. the linearly polarized flux density) has a deep
minimum. The measurement errors in Q and U are generally independent,
and the actual track of the Stokes vector would be a jittery version of
loop 2 in Figure~\ref{Fig2}b. A random small change in Q and/or U near
the time of PF minimum could throw the loop across the origin, change
the winding number, and so change the net EVPA rotation from 180\degr~to
0\degr. The effect can be described with Figure~\ref{Fig2}c, where $\rm
EVPA_2$ has a $\sim 90\degr$ CCW swing at the time of minimum flux,
and a net EVPA change of 180\degr. If noise were to throw the loop across
the origin then the EVPA curve would be similar except that the 90\degr\
swing at minimum flux would be in the CW direction and the net change
would be 0\degr.

From these examples it appears likely that small fluctuations in the
signal, or small variations in Q and U due to noise, can cause large
changes in the EVPA.  These effects would appear when the loop is
close to the origin of the (Q,U) plane; that is, when the amplitude
is in a deep minimum.

%%%%%%%%%%%%%%%%%%%%%%%%%%%%%%%%%%%%%%%%%%%%   SECTION 6
%Section 7
\section{Physical Models}
\label{s:physicalmodels}

In this Section we first consider the shock-in-jet calculations by
\citet{ZCB14, ZCB15} and then the MHD calculations by \citet{NGM10}
and \citet{NM14}. These provide some theoretical support for our
two-component model.

\citet{ZCB14, ZCB15} calculate
the radiation from a transverse shock in a relativistic cylindrical
plasma jet in a helical magnetic field.  For our purposes it can be
regarded, in essence, as a two-component model with the background jet
being the first component, and the passage of the shock producing a burst
that is the second component.  Integration to find the radiation from the
shock is in elementary diagonal disks that allow for the light-travel
time across the jet, such that all photons from a disk are received at
the same time by the observer. This means that a burst starts at zero
amplitude, builds up to some peak amplitude that may persist for a while,
and then symmetrically goes to zero. This is closely analogous to our
parabola model. The parameters in the calculation can be picked to give
the same results as we obtained; for example, the double 90\degr~step
in \citet[][Figure 7]{ZCB14} is the same as the double 90\degr~step in
Figure~\ref{Fig4} above.

The geometry of the system automatically makes the radiation from the jet
and the burst orthogonally polarized, if the pitch angle of the helical
field, and the angle to the line-of-sight, are in the appropriate ranges.
This has been discussed in detail by \citet{LPG05}. The shock-in-jet
model of course is itself an idealized case, with perfect cylindrical
symmetry, but it does provide a physical model for the two-component
model as we have used it.

In our two-component model the net EVPA rotation can be 180\degr~if
the EVPA rotation in the burst itself; i.e.,  the quantity $\Phi$ in
Figure~\ref{Fig1}, is small, provided the EVPA difference between the
jet and the burst is close to 90\degr.  In a realistic situation $\Phi$
can be small, but not zero; i.e., the burst must have some rotation.
Where does this rotation come from? One possible answer lies in
the work of \citet{Nak01} and \citet{NGM10}, who studied a shock
in a relativistic jet that is threaded by a helical magnetic field.
They showed that the shock compresses the toroidal component of the
field, and the resulting increase of angular momentum in the field is
balanced by a counter-rotation in the plasma.  The passage of the shock,
with its rotating plasma, will produce a burst of emission with a rotating
EVPA. This burst emission adds to the jet emission and, if the conditions
are right, the observed net rotation can be 180\degr.  In addition, if
the jet speed is higher than the speed of the fast magnetosonic wave,
then the reverse shock will be carried forward in the galaxy frame, and
its radiation also will have an EVPA rotation, but in the opposite sense
from that of the original forward shock \citep{NM14}. This mechanism
was suggested as a means for generating the rotations with reversals
in successive bursts in OJ\,287 \citep{CAA18}.  In the next section,
analyzing more recent data on OJ\,287, the rotations are all CCW, and the
simpler version, with the jet speed faster than the slow magnetosonic
wave but slower than the fast magnetosonic wave, might be responsible
for $\Phi$, the rotation in the burst.

%%%%%%%%%%%%%%%%%%%%%%%%%%%%%%%%%%%%%%%%%%%%
%Section 8 
\section{OJ\,287} 
\label{s:oj287}

\begin{figure}[!ht]                   %% FIGURE 8 
\centering 
\includegraphics[trim={0 0cm 0cm 0}, clip, width=1.0\columnwidth]{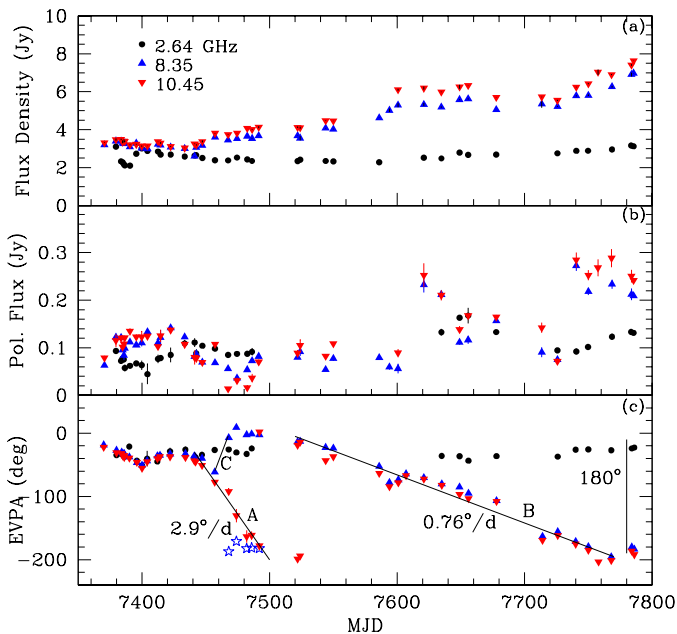} 
\caption { 
Data from \citet{MKA18} for OJ\,287. (a) total flux density (Stokes I)
(b) linearly polarized flux density (c) EVPA. The 180\degr~step at MJD
7492 in (c), at 10.45 GHz, is introduced to show that
the EVPAs at 8.35 and 10.45 GHz are closely similar after that date.
The five stars are 180\degr~below the corresponding 8.35 GHz
triangles and are discussed in the text.  Events A, B, and C are
described in the text. MJD 7400 corresponds to 12 January 2016.
} 
\label{Fig8} 
\end{figure}

In this Section we illustrate the use of our model with high-cadence
polarization observations of OJ\,287 published by \citet{MKA18}. We
obtained the relevant data from the CDS archive, and repeat part of it
here in Figure~\ref{Fig8}; namely, Stokes I $(\equiv \rm{F}$) and EVPA
at 2.64, 8.35, and 10.45 GHz (panels a and c), and the linearly polarized
flux density PF (panel b). PF is found by multiplying F by $m_l$ ($\equiv
\rm{P}$), the fractional linear polarization, which are in the CDS archive.
Some of the epochs for I and $m_l$ are not identical, and we only used
data for which the epochs differ by $0.2d$ or less.  Error bars for F
and EVPA are in the CDS archive, and the error bars for PF are found with
standard propagation of errors. In all three panels of Figure~\ref{Fig8}
most of the error bars are smaller than the points.

The EVPA plot in Figure 8c differs from the Myserlis et al plot (their
Figure 1c) in that we have introduced a $+180\degr$ jump at 10.45
GHz, at MJD 7492, to make it easier to see that the 8.35 GHz and
10.45 GHz points are closely similar after that date. Three 10.45
GHz points near MJD7500 are duplicated with a separation of 180\degr, to
help in following the change in slope, and we added a vertical bar of
length 180\degr~at MJD 7780, to emphasize that the EVPA is nearly the
same at all three frequencies at the end of the data run. In addition,
we corrected all the EVPA values for Galactic Faraday rotation, using
$\rm RM=+31.2~rad~m^{-2}$ \citep{TSS09}.

\begin{figure}[!ht]  %% FIGURE 9 
\centering
\includegraphics[trim={0 8cm 0 0}, clip=true, width=1.0\columnwidth]{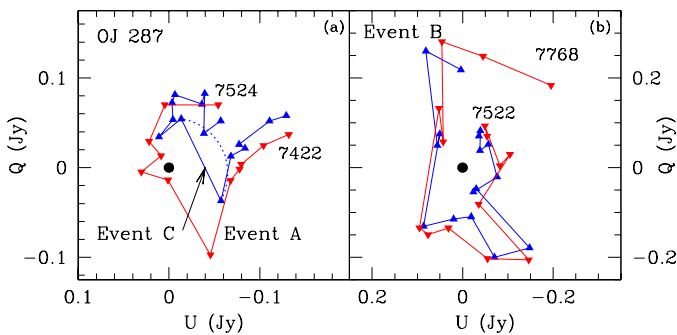}
%%% trim={left, bottom, right, top}
\caption{
Stokes plane showing the evolution of Q and U at 8.35 GHz (blue) 
and 10.45 GHz (red).
(a) Events A and C, running  from MJD 7422 to MJD 7524. Note that the
10.45 GHz loop (Event A) encloses the origin while the 8.35 GHz loop
(Event C) does not. The dotted arc corresponds to the straight-line
segments in Figures~\ref{Fig8}b and \ref{Fig8}c; see text.  (b) Event B,
running from MJD 7522 to MJD 7768.  Note that the scales in (a) and
(b) are different.  
}
\label{Fig9}
\end{figure}

The short line at MJD 7460 connects points at 8.35 GHz that are
54\degr~apart. This is the customary connection, whereby $n\pi$
is added to an EVPA value to make adjacent points differ by less than
90\degr.  This is the connection  that was used by \citet{MKA18} and is
seen in their Figure 1.  With it we see that the behavior of the EVPA
in the interval MJD 7440--7500 is different at 8.35 GHz and 10.45 GHz.
At the higher frequency there is a rotation of at least 140\degr~CW,
while at the lower frequency the rotation is about 70\degr~CCW.  This
is surprising since the frequencies are only 0.1 dex apart, and the
spectrum is rather flat, with index $\alpha \sim 0.5$ \citep{MKA18}.

An alternative connection for the 8.45 GHz points subtracts
180\degr~at MJD 7472 and later, giving a jump of $-126\degr$
CW.  The first five of these points are shown with stars in
Figure~\ref{Fig8}c. There still is a substantial difference between the
curves at the two frequencies, since the 8.35 GHz curve now has an abrupt
drop with a slope of at least $-4.9\degr/d$, while the 10.45 GHz curve
has a steady drop of $-2.9\degr/d$.  Hence, with either connection, we
conclude that the EVPA curves show a substantial difference between 8.35
and 10.45 GHz. This is discussed further in Section~\ref{s:stokes}. In
the rest of this discussion we use the first connection, consisting of
the blue triangles in Figure 8c.

The two rotations seen at 10.45 GHz have different slopes as shown in
Figure 8c, and we regard them as comprising two distinct events, called
A and B.  Event B also appears in the 8.35 GHz data, and the polarized
flux density and EVPA are closely similar at the two frequencies. The
weak Event C appears at 8.35 GHz in the middle of Event A. These 
events are not seen at 2.64 GHz. Presumably, they occur deep in the
core and are hidden by a large optical depth at 2.64 GHz.

The polarized flux densities in Figure 8b have deep minima at the time
of Events A and C. This is not a coincidence but is a feature of our
two-component model, described in Section~\ref{s:2-comp}, and is due to
the (Q,U) loop coming close to the origin.  We now describe the details
of these events.

In Event C the 8.35 GHz points at first follow the 10.45 GHz points,
then at MJD 7460 have a CCW rotation of about +75\degr. After the gap
at MJD 7500 the 8.35 and 10.45 points lie close together and have a
slow CW rotation of about 180\degr.  The earlier rotation at 10.45 GHz
follows a reverse S-shaped curve, with wings that slowly approach the
base level, at least on the early side.  The line marked $2.9\degr/d$
shows that the rotation is nearly uniform in its central region; this
line is not a fit to the data but was drawn by eye.

The 10.45 GHz points at MJD 7492 and 7520 are duplicated with a
separation of 180\degr, and probably belong to both events A and B. In
our models (Section~\ref{s:2-comp}) the burst begins and ends at low
amplitude, and so the net EVPA rotation must be gradual at the beginning
and end of the event, as it is for Event A.

The slow rotation in Event B is the same at 8.35 and 10.45 GHz, and is not
seen at 2.64 GHz.  The line marked $0.76\degr/d$ is drawn by eye and shows
that the rotation has small but significant departures from being uniform.

%% SECTION 8.1
\subsection{Stokes Plane} 
\label{s:stokes}

The data in Figures~\ref{Fig8}b and \ref{Fig8}c are shown on the
Stokes plane in Figure~\ref{Fig9}.  The fast EVPA rotation,
defined somewhat arbitrarily in the interval MJD 7422 -- MJD 7524,
is in Figure~\ref{Fig9}a.  We have connected the successive (Q,U)
points with straight lines, although these do not match the corresponding
straight line segments in Figures~\ref{Fig8}b and \ref{Fig8}c. In
Figure 9a the dotted arc corresponds to the straight segment C in Figure
8c, and its radius varies only a little because the corresponding
PF endpoints are close together.

In Figure 9a the loops at the two frequencies are similar, except that
one encloses the origin and the other does not. Thus the two EVPA
curves are different. This appears to be rather accidental, and
perhaps due to fluctuations or simply to spectral differences in the
polarized flux densities of the jet and the burst. If the amplitude of
the 10.45 GHz burst had been 15\% smaller, or the one at 8.35 GHz 20\%
larger, while the jet remained the same, then the two curves would
have had the same rotation.

In Figure~\ref{Fig9}a the loops are roughly symmetric about the line from
the origin to the apex of the loop. This means that, in the two-component
model, the jet and the burst must be nearly orthogonally polarized. This
is the case at both frequencies; i.e., for both Events A and C.

In Figure 9b the loop for Event B is essentially frequency-independent and
is much slower that the fast loop in Figure 9a. Event B itself may
be a multiple event with two major excursions in amplitude, seen at
MJD 7620 and MJD 7740 in Figure 8b. Note that the scales in Figures 8a
and 8b are different.

\begin{figure}[ht]  %% FIGURE 10 
\centering 
\includegraphics[trim={0cm 0cm 1cm 0.5cm}, clip, width=1.0\columnwidth]{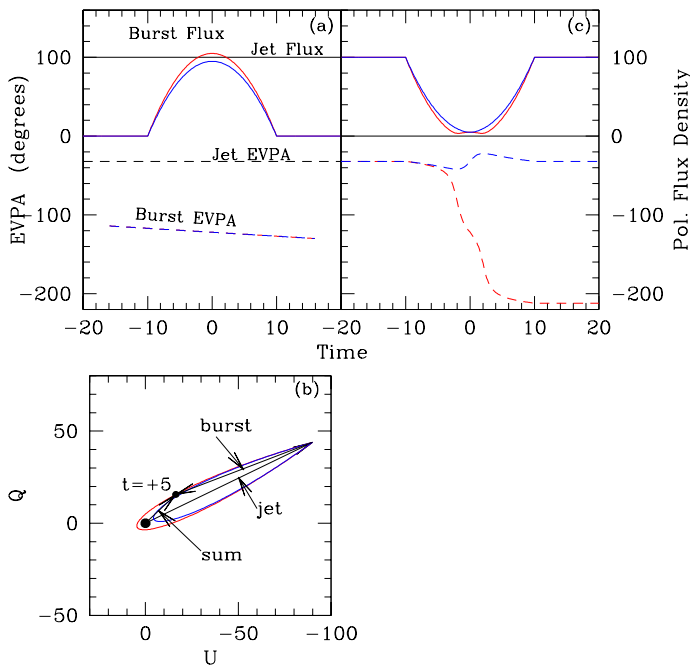}
\caption{ Two-component models (Table 2) that have the same topology
  on the (Q,U) plane as the fast rotations in Figure 9a. The model for
  8.35 GHz is blue, and that for 10.45 GHz is red.  (a) as in
  Figure~\ref{Fig1}.  (b) QU plane for the models in (a). The steady
  jet is shown as the long vector from the origin, the loops show the
  tracks of the sum (jet plus burst) vectors. The burst and sum
  vectors are shown at $t=+5$.  (c) The flux density and EVPA of
    the sum. See text.  }
\label{Fig10}
\end{figure}

\begin{table}
\begin{center}
\caption{Parameters for the models in Figure~\ref{Fig10}}
\label{t:models} 
\begin{tabular}{lccccl}
\hline\hline
Frequency  &    R      & $\Phi$ & $\Psi$ \\
   GHz     &           &  (degr) &  (degr) \\ \hline
   8.35    &  0.95     &   10   &   90   \\
   10.45   &  1.05     &   10   &   90   \\
\hline
\end{tabular}
\end{center}
\end{table}

%% SECTION 8.2
\subsection{Two-Component Models}
\label{s:oj287models}

Figure 10a shows models for Events A and C in Figure 9a.  Details of
the models are in Table~\ref{t:models}. The jet is the same for the two
frequencies and is constant in time. R is the ratio of the peak
polarized flux of the parabola to that of the jet.  The two ratios are
roughly related by the spectral index of OJ\,287, which is near 0.5. The
angles $\Phi$ and $\Psi$ are shown in Figure 1; $\Phi$ is the total phase
change across the burst, and $\Psi$ is the EVPA difference between the
jet and the burst, at $t=0$.

Figure 10b shows the models on the Stokes plane. Like the data (Figure
9), the 10.45 GHz loop (red) encloses the origin but the other one
(blue) does not.  However they both come close to the origin on the
(Q,U) plane and this makes the two model fluxes small at that time.
In Figure 10c, showing the model observables, the EVPA is
symmetric around $t=0$, because we took $\Phi = 90\degr$. This symmetry
can be seen in the data at 10.45 GHz, around MJD 7470.
At 8.35 GHz the model EVPA
(Figure 10c, blue) goes through an S-shaped excursion of amplitude 20\degr;
the data go through a similar but larger excursion of 79\degr~between
MJD 7457 and MJD 7474. This is a large difference between the data and
the model, and is discussed below.

The nature of the minima in the model amplitudes also matches those in
the data. In Figure 10c, at 10.35 GHz, the minimum consists of two low
points with a slightly higher value in the middle. At 8.35 GHz there
is a single minimum accurately located in the center of the 10.45 GHz
minimum. These both match the data in Figure~\ref{Fig8}b.

To produce all these similarities, the model parameters must be rather
close to those in Table~\ref{t:models}. We did not attempt to fit the
models to the data, but did pick the parameters to bring the models close
to some of the features in the data. Thus the jet and the peak of the
burst must be similar in amplitude and close to orthogonal because the
measured PF values are very low, and time-symmetric.  The value
of $\Phi$ is not well-fixed. We chose $\Phi$, rather arbitrarily, as
10\degr, for both frequencies. This gives an 8-GHz S-shaped excursion
around $t=0$ with an amplitude of about 20\degr, while the data have
a similar excursion of 79\degr~between MJD 7457 and MJD.7474. Making
$\Phi_{8.35}$ larger would increase the amplitude of the excursion and
make the fit between the model and the data better; however, increasing
$\Phi_{10.45}$ would change the nature of the flux minimum, turning the
double dip into a single dip. We could choose different values of $\Phi$
at the two frequencies, but by the argument we used above they should
be close together because the frequencies differ by only 0.1 dex.

This discrepancy in Event C at 8.35 GHz is the largest difference between
the model results and the data.  We suspect that it has to do with the
poor match between the shape of the data loop in Figure~\ref{Fig9}a and
the shape of the sum loop in the model, Figure~\ref{Fig10}.  We used a
parabola to simulate the amplitude of the burst, for simplicity. This
gives a model that matches the data in several ways, but its smoothness
is unlike that in real bursts, which usually are spiky and not symmetric
around the peak.  A better fit could no doubt be found by adding more
parameters to the model.  However, a detailed development of a model,
together with a non-linear fitting procedure, does not seem warranted
at this time.

\citet{MKA18} regard the large EVPA rotations seen in
Figure~\ref{Fig8} as tracing motions of the emission region around
a bend in the jet, or on a helical trajectory. We suggest, as an
alternative, that they primarily are phase effects, as in the models in
Figure~\ref{Fig10}.  There are two reasons for this. First is the good
fit of the data to the model, which has a physical rotation $(\Phi)$
of only 10\degr. Admittedly, the value of $\Phi$ is not well fixed, but
it appears to be much smaller than 180\degr.

A second reason to think that the model is correct and that the EVPA
rotation is mainly a phase effect is the strong difference in the EVPA
curves between 8.35 and 10.45 GHz. This difference is unlikely to be
due to optical depth effects and is not readily explained in terms of
helical or other motions. We ascribe it to a minor difference in the
details of the bursts at the two frequencies. The burst and the jet
are nearly orthogonal and have nearly equal values of PF, but the
differences are enough to make the winding number at the two frequencies
different.  We suggest that the model does describe the real situation. It
generates 180\degr~rotations similar to those observed, and the required
near-orthogonality is actually seen in the data \citep{MKA18}.

%%%%%%%%%%%%%%%%%%%%%%%%    SECTION 8  DISCUSSION
% Section 9
\section{Discussion}
\label{s:discussion}

Our main result is the simple statement in Section~\ref{s:simple},
that the EVPA rotation across the combination of a burst with a steady
jet is $n\pi$, where n is an integer.  The first consequence of this
is that small physical rotations of the source itself can lead to large
EVPA rotations; for this to occur the background and the burst must have
EVPAs that are close to orthogonal. The possibility of small physical
rotations leading to large EVPA rotations was also noted in an earlier
paper \citep{CAA18}.

The second consequence is that special behaviors are possible,
especially in more realistic models where the the burst is not
smooth.  In Figure~\ref{Fig5} the EVPA rotation is 360\degr~CCW; and
in Figure~\ref{Fig6}, using the same parameters except for a small
change in the amplitude of the burst, the rotation is 180\degr~CW plus
a substantial overshoot. In the latter case the EVPA rotation is in the
opposite sense to the rotation of the sub-components. This drastic change
comes about because the topology of the loop on the Stokes plane changes;
the winding number changes from $+2$ to $-1$.

Another consequence is that small fluctuations can have large effects on
the EVPA. In Figure~\ref{Fig7}, 6\degr fluctuations
cause a rotation of 180\degr, and the rotation can be in either direction,
depending on the relative phase and amplitude of the jet and the burst.
In Section~\ref{s:oj287models} we suggest that this process is responsible
for the $\sim 180\degr$ EVPA rotation seen at 10.45 GHz but not at 8.35 GHz
in OJ\,287 \citep{MKA18}.  The fluctuations can also be due to internal
noise and measurement error, which can change the winding number if 
PF is near a minimum.  These are different processes
from previous discussions of stochastic methods for generating large
EVPA rotations, which use a set of randomly polarized emission cells
\citep[e.g.,][]{KB17}. The large rotations occur in the random walk of
the the net EVPA.

In our analyses, except for the sinusoidal ripple in
Section~\ref{s:noise}, the burst has been smooth and its fall has been the
reflection of its rise. But this is not realistic, and observed bursts
in polarized flux are usually irregular in both EVPA and amplitude.
This means that loops on the Stokes plane will be irregular, not
smooth and symmetric as in Figures~\ref{Fig2}, \ref{Fig4}, \ref{Fig5},
and \ref{Fig6}.  We see this irregularity in Figure~\ref{Fig9},
which shows the Stokes plane for OJ\,287.  Other (Q,U) plots in the
literature are similarly irregular; see e.g., \citet[][Figure 16]{V10}
and \citet[][Figure 11]{CAA18}.

%%%%   SUBSECTION 9.1
\subsection{Simultaneous Rotations at Radio and Optical Wavelengths}
\label{s:radioopt}

The literature reports several cases where large EVPA rotations are
closely similar at different wavebands; e.g., at radio and optical or
IR frequencies \citep[e.g.,][]{K88, DM09}. and this has been taken
as an indication that the source region is the same at both wavebands.
In this case the phase mechanism we have described might be at work,
with the following scenario.

If the source region is the same for the two bands, then their geometries
are similar. For the core of OJ\,287 at 43 GHZ, \citet{DM09} showed that
the EVPAs of the background jet and the new component (presumably, a shock
that generates the burst) were either parallel to or perpendicular to the
position angle of the jet. This would apply at both wavebands and would
give the opportunity for the jet and the burst to be orthogonal.  The jet
EVPA would be the same at both bands, and so the two rotations would
start and end at the same EVPA.  However, the details of the rotation
would also be affected by the shape and amplitude of the bursts due to
the shock, and further study is needed to establish if these could be
similar at the two frequencies.

%   SECTION 10
\section{Summary and Conclusions}
\label{s:summary}

We investigate a simple two-component model for generating large EVPA
rotations in the emission from a blazar. The model consists of a steady
``jet'' and a variable ``burst'', although by ``steady'' we merely
mean that this component has a much longer time constant than the more
variable burst. These two components are represented by vectors on the
Stokes (Q,U) plane, where the jet vector is fixed and the burst vector
rotates. The sum of the two vectors forms a loop on the (Q,U) plane,
and is the observable quantity.  The topology of the loop controls
the EVPA swing, and the net EVPA rotation across the burst is $n\pi$
where $n$ is the winding number of the loop (number of times it encloses
the origin.)  Three parameters control the details of the rotation: R,
the ratio of the peak polarized flux of the burst to that of the
jet; $\Phi$, the overall EVPA swing of the burst; and $\Psi$, the EVPA
difference between the jet and the burst at its peak.  The allowed
combinations of R, $\Phi$, and $\Psi$ for the net rotation of EVPA
to be 180\degr~rather than 0\degr~are shown in Figure~\ref{Fig3}.
The possibility for a 180\degr~ swing always exists for $R>1$ provided
the point is above the critical line in Figure~\ref{Fig3}, but if
$\Phi$ is small then $\Psi$ must be close to 90\degr; i.e., the EVPAs
for the jet and burst must be nearly perpendicular. This means that a
180\degr~swing can be generated with little EVPA rotation in the burst
itself. This is a phase effect and does not require the physical
rotation in the plasma to be comparable to that in the observed EVPA.

The model can accomodate a wide variety of behaviors, many of which
are seen in the observations.  If $\Phi$ is small then the (Q,U) loop
is thin, and the rotation can consist of two separated 90\degr~swings,
which can be of the same or opposite sense, giving a net rotation of
180\degr~or 0\degr~(see Figure~\ref{Fig4}). If the burst has structure
as in Figure~\ref{Fig5}, then the EVPA swing is stepped and, since in
this case the winding number is 2, the net swing across the burst is
360\degr. In Figure~\ref{Fig6} the parameters are carefully chosen, but
this example shows that the observed EVPA rotation can be in the opposite
sense from that of the burst itself. Note also that in Figure~\ref{Fig6}
the EVPA has an overshoot; the overall swing is 237\degr~although the
net swing across the burst is 180\degr.

Small fluctuations in the burst, or system noise, can provide variations
in $\rm R, \Phi~or~\Psi$ that cause 180\degr~rotations. For this to
occur R must be near unity and $\Psi$ must be near 90\degr.

The use of the model is illustrated with recent data from OJ\,287 at
2.64, 8.35, and 10.45 GHz \citep{MKA18}. At 10.45 GHz there are two
EVPA rotations of order 180\degr. Although they are adjacent in time,
we regard them as different events because the rates are different, and
because one of them appears only at 10.45 GHz.  We suggest that
the difference in the EVPA behavior between 8.35 GHz and 10.45 GHz is
accidental, and is due to small differences in the bursts (or the jet)
at the two frequencies such that the winding numbers are different.

We briefly consider the observations of simultaneous EVPA rotations
at radio and IR or optical bands, and show how our model provides
a plausible mechanism for this phenomenon.

The shock-in-jet model of \citet{ZCB14, ZCB15} has two emission
components, the quiescent jet and the transverse shock.  Integration
across the shocked region that keeps track of the light-travel
time gives a symmetric burst of emission, similar to our parabolic
burst. Results for the EVPA rotation can be closely similar to our results
in Section~\ref{s:2-comp}. The required orthogonality of 
the two components is not an ad-hoc assumption but is a result of the
helical nature of the magnetic field and the symmetry of the system.

We conclude that the two-component model can explain a wide variety
of observed EVPA rotations. Under certain conditions, the rotations
can be phase effects, with little connection to physical rotations in
the plasma. 

%%%%%%%%%%%%%%%%%%%%%%%%%%%%%%%%%%%%%%%%%%%%%%%%%%%%%%

\begin{acknowledgements}

We are grateful to D.~Blinov and the RoboPol collaboration for giving us
data in advance of its publication, to T.~Hovatta for her comments
and help in arranging the data transfer, and to the referee, who
made a number of comments that improved the manuscript. MHC thanks
S.~Kiehlmann, D.~Meier and T.~Pearson for discussions at an early stage
of this work, and G.~Jennings for a discussion of the winding number.
We are grateful to I.~Myserlis for his careful reading of an early version
of the manuscript, and to M.~Lister and Y.Y.~Kovalev for comments.  TS was
supported by the Academy of Finland projects 274477, 284495, and 312496.

\end{acknowledgements}

%REFS
%

%APPENDIX

%% APPENDIX A
\begin{appendix}
\section{Loops on the Stokes Plane}
\label{s:loops}

Loop 2 in Figure~\ref{Fig2}b comes close to the origin of the (Q,U)
plane, and the character of the associated EVPA curve changes if the axis
of the loop rotates enough to change the winding number. In this
Appendix we calculate the conditions for the loop to touch the origin.

We first find the time, $t_c$, at which the loop touches the origin. 
The Stokes parameters are additive and so we write 
$\rm Q_{sum} = Q_{jet} + Q_{burst}$ and $\rm U_{sum} = U_{jet} + U_{burst}$.
When the loop touches the origin, $\rm Q_{sum}=0$ and $\rm U_{sum}=0$, or
\begin{equation}
  {\rm A \cos(\xi_{jet}) = -AR[1-({\it t_c} /w)^2] \cos(\xi_{jet} - 2\Psi + \Phi({\it t_c} /w))}
\end{equation}
\begin{equation}
  {\rm A \sin(\xi_{jet}) = -AR[1-({\it t_c} /w)^2] \sin(\xi_{jet} - 2\Psi + \Phi({\it t_c} /w))}
\end{equation}
where A is the flux density of the jet, AR is the peak flux density of
the burst, and $\Psi$ and $\Phi$ are defined in Figure~\ref{Fig1}.
Squaring and adding gives ${\rm R^2[1-({\it t_c} /w)^2]^2 = 1}$ and 
\begin{equation} 
   {t_c = {\rm \pm w \sqrt{1-1/R}}} 
\label{t_c} 
\end{equation} 
We choose the minus sign in the square root because $(t_c/{\rm w})\le 1$.

At $t=t_c$ we can write
\begin{equation}
   {\rm EVPA_{burst,c} = EVPA_{jet} - \Psi + \Phi({\it t_c} /2w)}
\end{equation}
\begin{equation}
   {\rm    = EVPA_{jet} - \Psi \pm (1/2)\Phi \sqrt{1-1/R}}
\end{equation}
But at $t=t_c$, on the (Q,U) plane, the burst and jet vectors
are opposite and cancel; hence 
${\rm EVPA_{burst,c} = EVPA_{jet} \pm 90\degr}$ and
\begin{equation}\label{PsiPhi}
   {\rm \delta = \pm (1/2)\Phi \sqrt{1-1/R}}
\end{equation}
where $\delta = \Psi - 90\degr$ i.e., $\delta$ is the departure of
the jet and burst from orthogonality, at $t=0$. This may be written as
\begin{equation}\label{Rdelta}
   {\rm R = [1-(2\delta/\Phi)^2]^{-1}}
\end{equation}
Equation~\eqref{Rdelta} is plotted in Figure~\ref{Fig3} as the critical
line.  Points on the line have (Q, U) loops that touch the origin. Points
above the line have an EVPA rotation of 180\degr, while those below the
line have a net rotation of 0\degr.

\end{appendix}

\end{document}